# Time-Variant Proof-of-Work Using Error-Correction Codes

Sangjun Park, Haeung Choi, and *Heung-No Lee, *Senior Member, IEEE*

*Abstract—* The protocol for cryptocurrencies can be divided into three parts, namely consensus, wallet, and networking overlay. The aim of the consensus part is to bring trustless rational peer-to-peer nodes to an agreement to the current status of the blockchain. The status must be updated through valid transactions. A proof-of-work (PoW) based consensus mechanism has been proven to be secure and robust owing to its simple rule and has served as a firm foundation for cryptocurrencies such as Bitcoin and Ethereum. Specialized mining devices have emerged, as rational miners aim to maximize profit, and caused two problems: *i)* the re-centralization of a mining market and *ii)* the huge energy spending in mining. In this paper, we aim to propose a new PoW called Error-Correction Codes PoW (ECCPoW) where the error-correction codes and their decoder can be utilized for PoW. In ECCPoW, puzzles can be intentionally generated to vary from block to block, leading to a time-variant puzzle generation mechanism. This mechanism is useful in repressing the emergence of the specialized mining devices. It can serve as a solution to the two problems of recentralization and energy spending.

## I. Introduction

In cryptocurrencies, the consensus part plays a role in leading an agreement among trustless nodes without any communications. This part is the most innovative because it can prevent the double spending attack [1] in a peer-to-peer network in the absence of trusted parties. In *Bitcoin* [2], as an example, more than ten thousand of nodes randomly scattered across the world aim to reach a consensus in each block time. The Internet is the only way to connect them; communication packets are delayed and sometimes dropped though the Internet that is designed to provide the best effort service. Cyberattacks frequently happen, making transactions over the Internet insecure. Nevertheless, Bitcoin has shown secure peer-to-peer transactions over the past 10 years. With the help of proof-of-work (*PoW*) which is fundamental to the consensus part, this can be possible.

In Bitcoin, each node does competitive work, called *mining*, to forge a block. The node which wins this competition has the right to mint a specified number of coins as this mining reward. If a node was re-forging all the blocks alone, it could spend the total amount of works done to all the mined blocks.

Without PoW, anybody with a computer can alter the content of the blockchain, implying unauthorized changes in any mined blocks can be possible. If PoW is attached to each mined block, attackers cannot make any unauthorized modifications without redoing all the works. No node can alone alter any mined block, meaning an immutability property.

In Bitcoin, miners make rational decisions to maximize their profits by following a two stage process in which *i)* the miners select a blockchain whose length is the longest and *ii)* they extend this longest one by adding a newly mined block. Suppose there are two blockchains where one is longer than the other one in terms of the length. Since the longer chain has the more accumulated works, altering it is more difficult. This longer chain shall be treated the more trustable and preferable by the miners. Thus, they select the longer chain. Making such a selection is rational for the sake of keeping the mining rewards. The mining reward is a delayed conditional payment, i.e., if a miner mines a block at a given time point $t_1$, the reward is delayed until the future moment $t_2$ of time. This time from $t_1$ to $t_2$ is measured in terms of the number of blocks, say 100 blocks. If this mined block was not a part of the longest chain at the future time point $t_2$, the reward vanishes. Thus, rational miners select the longest chain.

In Bitcoin, miners spend computational resources to forge a block by solving a puzzle carved in a bitcoin program as an on-chain policy. This puzzle is made using the secure hash algorithm 256 (SHA256) [3]. To solve the puzzle, the miners have to repeatedly execute SHA256 by varying an input to SHA256 until a good hash is given. This input is the header of the block, i.e., block header, including six fields such as version, previous hash, difficulty, timestamp and nonce, Merkle tree value. The version is fixed. Given a block at a certain height, i.e., the $l^{th}$ block, the previous hash and the difficulty are obtained from its previous block, i.e., $(l–1)^{th}$ block. They are constant. The rest varies until SHA256 returns a good hash. A good hash can be spotted since it shall possess a certain number of leading zero bits reflecting the difficulty. The block header of a mined block can serve as the proof that a given puzzle is solved without any falsehood.

Satoshi [2] intended for miners to execute SHA256 using a central processing unit (CPU). But, faster computing machines based on application-specific integrated circuit (ASIC) became available. As a result, the miners have chosen to exploit them to maximize their profits. To date, the miners equipped with ASIC

S. Park is with the Electronics and Telecommunications Research Institute (ETRI), South Korea (e-mail: sjpark86@etri.re.kr).

H. Choi and H.-N. Lee are with the School of Electrical Engineering and Computer Science (EECS), Gwangju Institute of Science and Technology (GIST), South Korea (e-mail: haeung@gist.ac.kr, heungno@gist.ac.kr).

This work was supported in part by the National Research Foundation of Korea (NRF) Grant funded by the Korean government (MSIP) (NRF-2018R1A2A1A19018665). The corresponding authors: Heung-No Lee (heungno@gist.ac.kr).

mining devices have dominated the mining business [4], leading to two problems:

M1. The mining markets have become re-centralized [5].
M2. The electrical energy spent to mine blocks is huge [6].

First, the miners have a large portion of the total hash power, implying that the plight of the blockchain is left to a handful of these influential miners. It can be possible to modify any mined blocks on their own rights, leading to shattered trust. Namely, they can break the immutability property. Second, new models of ASIC mining devices can surpass old models with respect to the hash power, which is measured as a hash rate. Each miner is forced to buy newer models to win the mining competition. As the new models are widely used, the total hash power inevitably grows. The difficulty level to solve the puzzle must increase to keep a certain predefined range of block generation time that is expectedly consumed to mine a single block. Today, this level has gotten huge, i.e., $O(10^{20})$ hash rate per second. As a result, using CPUs in mining has gotten no longer profitable. Besides, as the total hash power increases, the miners spend more and more electrical energy to mine blocks.

If we prevent the usage of ASIC mining, we can alleviate the problems M1 and M2. To this end, we use the error-correction codes and their decoder. In general, the aim of using the codes in modern communication systems is to combat errors occurring over noisy channels in which the errors introduced over a noisy channel can be corrected by running a decoding algorithm. The codes have a rich history where there are numerous classes of good codes available. They have been used to define both a good public-key crypto system and a good hash function.

The first result can be traced back to the late 1970. McEliece [8] used Goppa codes to make a McEliece cryptosystem where a message is encoded using a public key $\mathbf{A} := \mathbf{SGP}$ where $\mathbf{G}$ is the generator matrix of a binary Goppa code, $\mathbf{S}$ is a nonsingular random matrix and $\mathbf{P}$ is a permutation matrix. The hash, i.e., the encoded result, of a provided message is made as follows: *a)* a word is made by multiplying the message with the public key and *b)* adding a binary random word whose number of ones is at most $t$ to this word[1] is to get the hash. Peters *et al.* [9] extended this system using non-binary Goppa codes to reduce the size of its public key. Even the size reduces, this system can achieve still the same security level as much as that of [8] could. Other codes such as low density generator matrix codes [10], low density parity check (LDPC) codes [11], [12], Reed-Solomon codes [13] and Reed-Muller codes [14] have been used to replace the Goppa codes. The aim of using them is to reduce the size of the public key.

Aside from the applications of the error-correction codes into the McEliece cryptosystem, the codes are used to construct new hash functions. Preneel [15] proposed a method to make new hash functions and proved that their hash functions can provide strong collision resistant properties. The codes in [15] are either the maximum distance separable codes or the Hamming codes.

Selman *et al.* [16] used LDPC codes to make a hash function and proved this function as an average universal hash function defined in [17]. These results can motivate us to exploit error-correction codes in designing a new PoW framework.

The contributions of this paper are three folds. First, we propose a new PoW framework which we name as ECCPoW. As the name implies, we add the error-correction code part into PoW. To the best of our knowledge, this is the first work in which the error-correction codes are applied to blockchains. We then explain how we make puzzles, which we call ECC puzzles, and give routines to solve them.

Second, we conduct a probabilistic study to examine a random variable called First Success Hash Cycle (FSHC) representing the number of hash cycles required in solving a given ECC puzzle. Based on this study, we get the following results:

- FSHC follows a geometric distribution with a parameter in (21) that depends on the number of miners $M$ and the code length $n$.
- The expected value of FSHC is a decreasing function with respect to the number of miners $M$.
- The expected value of FSHC is an increasing function with respect to the code length $n$.

Third, we define five properties for a good PoW and explain how ECCPoW satisfies these properties. We shall note that the most innovative property is the time-variant property, making ECCPoW suitable to resolve the problems of recentralization and energy spending.

We organize the rest of this paper as follows. Section II gives literature surveys regarding SHAs and PoWs. Section III elucidates LDPC codes and a decoder. Section IV addresses how ECCPoW works and gives its pseudo codes Section V presents theoretical results of ECCPoW. Section VI discusses properties of ECCPoW. Section VII presents the conclusions of this paper.

II. LITERATURE ON BLOCKCHAIN CRYPTOGRAPHIES

A. *Secure hash standard and functions*

The secure hash standard was formulated by NIST [3]. The purpose of this standard is to offer the specifications of SHAs that yield a hash of a given message. Even the message changes slightly, the hash of the changed message comes out completely different from that of the original message. Thus, a hash can be used to detect whether an original message was altered or not. SHAs with such a property can be used for the generation and the verification of digital signatures as well as for the message authentication.

A secure hash function takes an arbitrarily sized message and produces a fixed-size hash. Let a function $h$ be

$$h : \mathcal{X} \to \mathcal{Y}$$

which is said to be a cryptographically secure hash function if it satisfies the three requirements defined in [17] below:

(**One-way function**) Given any hash $y$ to which a corresponding message is not known, it is computationally infeasible to find a message $x$ such that $h(x) = y$.

---

[1] For any Goppa code, there is a construction method to guarantee that the minimum distance $d$ of that code is greater than a given positive integer. Thus, the value of $t$ can be known in advance using Theorem 1 [32].

TABLE I. The routine of bitcoin

Inputs: $\mathcal{S}_0$ (block header except for nonce) and $L$

| Step | |
|---|---|
| Step 1: | for nonce = 0, 1, 2, … $2^{32}-1$ |
| Step 2: | $\mathbf{e} = \text{SHA256}(\mathcal{S}_0 \cup \{\text{nonce}\})$ |
| Step 3: | If $\mathbf{e}$ begins with $L$ zero bits, then go to Step 5. |
| Step 4: | end |
| Step 5: | Block generation & broadcast |

(*Weak collision resistance*) Given an arbitrary message $x$, it is computationally infeasible to find any message $x'$ which results in the same hash, i.e., $h(x) = h(x')$.

(*Strong collision resistance*) It is computationally infeasible to find any two different messages $x$ and $x'$ which make the same hashes $h(x) = h(x')$.

NIST [3] has proposed a family of SHAs including SHA1, SHA224, SHA256, SHA384, and SHA512. A message of any size less than $2^{64}$ bits can be given as an input for SHA1, SHA224, and SHA256, while that of less than $2^{128}$ bits for SHA384 and SHA512. The size of a hash ranges from 160 to 512 bits, depending on the algorithm.

*B. PoW of Bitcoin*

In Table I, we define a puzzle in Bitcoin and give routines to solve this puzzle. In Step 2, a miner puts a given block header with a selected nonce to SHA256 and obtain a hash of 256-bits. In Step 3, this puzzle is declared to be solved if a hash is smaller than a specified target value, or in a simpler argument it begins with $L$ zero bits, where the target value or the value of $L$ is given as the difficulty level of the PoW puzzle. The miner repeats the routines from Step 2 to Step 3 by varying the nonce. However, there exists a chance in which the miner can fail to hit a good hash even though the whole range of nonces, i.e., $0 \sim 2^{32}-1$, is used. In such a case, the miner updates the block header, and repeats the routines from Step 2 and Step 3. There are two methods to update the block header. The first one is to update the timestamp field. The second one is to update the Merkle tree value by modifying the list of transactions being included in a block which the miner aims to construct.

*C. PoW of Ethereum*

Ethash [18] was created for the purpose of preventing the advent of ASIC mining devices in Ethereum. In Ethash, there is a memory structure called directed acyclic graph (DAG) where its data are randomly re-generated every 30,000 block.

Table II shows routines of Ethash. As we have shown in Step 2, the current block header with a nonce is taken by SHA3 to get a hash. This hash is taken by a predefined function to yield mix0 that is random. In Step 4, mix0 is used to determine which data from DAG are fetched. No one predicts which data shall be fetched from DAG because mix0 is random. The mixer takes both the fetched data and mix0 to get a random value in Step 5. In Step 6, mix0 is updated using the random value. The routines from Step 2 to Step 6 are repeated 63 times. Last, the decision is made using this final mix0, as we have shown in Step 8.

The ASIC resistant property in Ethash is originated from the fact that the operation time for the mixer is shorter than that of

TABLE II. The routine of Ethash

Inputs: $\mathcal{S}_0$ (block header except for nonce), $L$ and DAG

| Step | |
|---|---|
| Step 1: | for nonce = 0, 1, 2, … $2^{32}-1$ |
| Step 2: | $\text{mix0} = f(\text{SHA3}(\mathcal{S}_0 \cup \{\text{nonce}\}))$ |
| Step 3: | for $i = 1, 2, …, 63$ |
| Step 4: | $\text{data1} = \text{Fetch}(\text{DAG}, \text{mix0})$ |
| Step 5: | $\text{tmp} = \text{Mixing}(\text{mix0}, \text{data1})$ |
| Step 6: | $\text{mix0} = f(\text{tmp})$ |
| Step 7: | end |
| Step 8: | If mix0 begins with $L$ zero bits, then go to Step 10. |
| Step 9: | end |
| Step 10: | Block generation & broadcast |

where $f$ is a predefined function. Details on $f$ is given in [18].

the fetch operation. To be specific, let $A_i$ be the time duration (TD) to conduct the mixing operation in which the subscript $i$ denotes a chip to run this mixing operation. It is clear that the mixing operation TD based on an ASIC chip is significantly less than that based on any CPU chip because the clock speed of an ASIC chip is much faster than that of any CPU chip, i.e.,

$$A_{\text{ASIC}} \ll A_{\text{CPU}}.$$

Next, let $B_i$ be the TD to conduct the fetch operation. Unlike to the mixing operation TD, this TD depends on the communications bandwidth between the memory and the cache of the CPU in which the fetched data from DAG are passed through. In other words, the fetch operation time is connected mainly to the bandwidth but not to the clock speed. The purpose to use ASIC chips is to increase the processing speed; is not for obtaining a higher communications bandwidth. The fetch operation TD for CPU and ASIC are thus similar, i.e.,

$$B_{\text{ASIC}} \approx B_{\text{CPU}}.$$

We consider the inner routines of Ethash, i.e., Step 4 and Step 5. The mixing operation is conducted after the fetching operation is done. This operation TD at Step 4 can be significantly reduced using ASIC chips; but the fetching operation TD at Step 5 is not reduced even if ASCI chips are used. As a result, there is a bottleneck between Step 4 and Step 5. This bottleneck has the ASIC resistant property enabled.

Recently, a programmatic PoW (ProgPoW), which is planning to be used to replace Ethash, has been proposed to further improve the ASIC resistant property. This improvement is done by changing parameters related to DAG randomly from block to block. Such modifications can make the fetch operation time increased. But, the development of ProgPoW is not completed and ProgPoW is not proven to be secure at the time of writing this manuscript.

*D. PoW of Dash*

X11 was proposed in 2014 by Duffield [19]. In Table III, we give routines of X11 which consists of hash functions below:

Blake, Bmw, Groestl, Jh, Keccak, Skein, Luffa, Cubehash, Shavite, Simd and Echo.

TABLE III. The routine of X11

| | |
|---|---|
| Inputs: | $\mathcal{S}_0$ (block header except for nonce) and $L$ |
| Step 1: | for nonce = 0, 1, 2, ... $2^{32} - 1$ |
| Step 2: | $\mathbf{e} = \text{Blake}(\mathcal{S}_0 \cup \{\text{nonce}\})$ |
| Step 3: | $\mathbf{e} = \text{Bmw}(\mathbf{e})$ |
| | .... |
| Step 12: | $\mathbf{e} = \text{Echo}(\mathbf{e})$ |
| Step 13: | If $\mathbf{e}$ begins with $L$ zero bits, then go to Step 15. |
| Step 14: | end |
| Step 15: | Block generation & broadcast |

TABLE IV. The map for X16r

| Value | Hash | Value | Hash |
|---|---|---|---|
| 0 | Blake | 8 | Shavite |
| 1 | Bmw | 9 | Simd |
| 2 | Groestl | a | Echo |
| 3 | Jh | b | Hamsi |
| 4 | Keccak | c | Fugue |
| 5 | Skein | d | Shabal |
| 6 | Luffa | e | Whirlpool |
| 7 | Cubehash | f | Sha512 |

Blake first takes a given set of the current the block header with a selected nonce to get its hash. Next, Bmw takes this hash as its input to yield a hash. The same procedures are repeated until Echo, the last hash function, yields its hash. The decision is made using this last hash, as we have shown in Step 13.

However, the order of using the 11 hash functions is always fixed. This fixed order makes the development of ASIC mining devices possible and in fact an easy task. The development of ASIC mining solution can be done when these hash functions are implemented in a single device. Logic gates to sequentially connect the hash functions can be implemented. The first ASIC mining device targeting X11 was developed in 2016.

The idea behind X11 has been extended to other PoWs such as X13, X15, and X17. As the names suggest, they consists of 13, 15, and 17 hash functions, respectively. To date, there a set of ASIC mining devices for both X13 and X15 while there is no ASIC mining device yet for X17.

*E.  PoW of Raven*

In 2018, a new extension of X11 was proposed in [20]. This is called X16r. It uses multiple hash functions given in Table IV to get the last hash like to other extensions of X11 that we have discussed in Section II D. But, unlike the others, the sequence of the hash functions in X16r can be made to vary from block to block. This variation seems to be a role for preventing the advent of ASIC mining devices for X16r.

We provide an example to address how X16r operates. The sequence X16r is determined upon the last 16 bytes of a previous hash. Let this previous hash be

0x0000...04def2c3eff6da11542ffcdabce.

The last 16 bytes are 6da11542ffcdabce. Then, the sequence is decided on the basis of Table IV below:

Luffa → Shabal → Echo → Bmw → Bmw → Skein → Keccak → Groestl → Sha512 → Sha512 → Fugue → Shabal → Echo → Hamsi → Fugue → Whirlpool.

A miner puts a given set of the block header with a selected nonce through Luffa to yield a hash. Shabal takes this hash as its input to yield its hash. This routine is repeated until the last hash is yielded. In the above example, the last hash is obtained through Whirlpool.

At the time of writing this manuscript, it seems, no one has officially succeeded in implementing ASIC mining devices for X16r, i.e., there is no announced commercial product. However,

Black and Weight, the developers of X16r, in [20] stated that reordering the sequence cannot make the development of ASIC mining devices impossible. Recently, in [33] at Nov., 2019, Whitefire990 reported a simulation result which indicates the probability of $k$ time-repetition, such that the same hash function is used at least 5 times consecutively when $k$ is 5, exponentially decreases in $k$. Insisted further is that $k$ greater than 5 can be ignored in designing of ASIC mining devices. As such, what claimed there is that the ASIC-resistant property of Raven can be broken by the said ASIC designing method. It shall be noted, however, that all these claims provided in [33] have not yet been carefully verified through a peer-review system.

*F.  Short summary from C to E*

From the subsection II.C to II.E, we have reviewed the existing ASIC-resistant PoWs categorized as follows:

a. The usage of intentional memory access.
b. The usage of multiple hash functions.

Ethash and ProgPoW can belong to the first class while X11 and its variants such as X13, X15, X17 and X16r can belong to the second class. The basic idea of the first class is to use the bottleneck intentionally caused by randomly fetching data from a memory. The basic idea of the second one is to use the multiple hash functions, which can make the development costs of ASIC mining devices expensive.

At the time of writing this manuscript, ASIC mining devices for Ethash, X11, X13 and X15 are available. The development of ProgPoW is not yet available. X17 and X16r can be resistant to ASIC mining devices. But, as the ASIC-resistant property of the PoWs such as X11, X13 and X15 are broken, that of X17 can be cracked in the near future when the hardware development technology is improved. As mentioned in the subsection II.E, there is a claim that the anti-ASIC property of X16r could be broken; but to date no commercial ASIC mining device has been announced.

III.  Literature Surveys on LDPC

An LDPC decoder plays an important role in ECCPoW. That is, the decoder is utilized to generate an unpredictable random output that can be later on used to give a proof whether a puzzle is solved or not. We prepare this section to give a quick summary regarding the LDPC codes and their decoders.

In 1963, LDPC codes were proposed by Gallager in his thesis [7]. But, the codes received no attention back then because computers were not sufficiently fast enough to check the per-formance of a decoder. Mackay and Neal [22] reported in 1997 that the codes could achieve the Shannon limit [21] closely with a message passing decoder that uses a kind of belief propagation algorithms. Since then, numerous studies on the codes have been made. They are categorized as follows: *i)* constructing the codes to approach the Shannon limit [21] and *ii)* implementing the fast decoders based on either ASIC [23]–[26] or field programmable gate array (FPGA) [27]–[28] to support real-time decoding purpose.

*A. LDPC codes*

LDPC codes can be generalized to a non-binary alphabet for improving its error-correction capability. But, for the purpose of using these codes in ECCPoW to provide a new time-varying anti-ASIC PoW system, it hence is suffice to consider the binary alphabet version only.

An $(n, k)$ LDPC code is a linear code constructed by supplementing each message **m** of size $k$ with parity bits to get a codeword of size $n$. This code is often defined with respect to a parity check matrix **H** of size $m \times n$ such that each element is binary either zero or one and the number of ones is very small, where $m$ is the number of parity bits, i.e., $m = n - k$.

For a given parity check matrix **H**, its corresponding LDPC code can be either *regular* or *irregular*. If **H** contains a constant number $w_c$, called the column degree, of ones in each column and a constant number $w_r$, called the row degree, of 1s in each row, the code is called *regular*. For a given *regular* LDPC code, the parameters such as $n$, $k$, $w_c$, and $w_r$ satisfy the following:

$$nw_c = (n-k)w_r = mw_r. \quad (1)$$

If **H** contains a different number of 1s in both each column and each row, the code is called *irregular*. In the perspective of the error-correction capability, irregular codes are better than regular codes. To serve our purpose of anti-ASIC PoW mechanism, we aim to consider the *regular* LDPC codes because

i. it is much easier to implement a decoder of regular LDPC codes and
ii. the aim of using this decoder is not to correct errors but to yield an unpredictable random output.

A bipartite graph is often used to represent an LDPC code, as we have shown in Fig. 1. The lower and upper nodes are called the variable nodes and the check nodes, respectively. Each edge shows the adjacency of the $i^{th}$ variable node and the $j^{th}$ check node and corresponds to a nonzero $(i, j)^{th}$ element in **H**.

For a given LDPC code, its error-correction capability relies on the minimum (Hamming) distance $d$. This distance is given by solving an optimization problem in which we consider any pair of $2^k - 1$ different codewords below:

$$d = \min_{\mathbf{u} \in \{\mathbf{c}_1, \mathbf{c}_2, \cdots, \mathbf{c}_{2^k}\} \setminus \mathbf{0}_n} \|\mathbf{u}\|_h \quad (2)$$

which is NP-complete, where $\|\mathbf{x}\|_h$ is the number of 1s in **x**.

Thus far, studies on the computation of a good approximation to minimum distance for a given fixed **H** with reasonable size have been reported futile and as such it has remained as an open problem. Keha and Duman [29] proposed a branch and cut algorithm to obtain the minimum distance of LDPC codes. But, this algorithm requires a large amount of time and memory; it is thus only useful if $n$ is small. Then, Hashemi and Banihashemi [30] proposed a method to find lower and upper bounds of the minimum distance of LDPC codes and obtained both of the bounds even when $n > 64,000$.

For regular LDPC codes with a particular pair of $w_c$ and $w_r$, upper and lower bounds for a relative minimum distance which is the ratio of the minimum distance $d$ to the code length $n$, are given in [31] and [7], respectively. We use them for our purpose in this paper in Section IV. Once the minimum distance is given, the number of correctable errors can be computed as follows:

**Theorem 1 [32]**: Let a linear code be defined as a given parity check matrix **H** which has the minimum distance $d$. Then, the number of correctable errors is

$$t = \lfloor (d-1)/2 \rfloor \quad (3)$$

where $\lfloor x \rfloor$ denotes the integer part of $x$.

We explain how to encode a message **m** for a given **H** of size $m \times n$. To this end, we build a generator matrix **G** of size $n \times k$ whose column space is orthogonal to the row space of **H** below:

Step 1: Conduct the Gaussian elimination to rewrite **H** as follows:

$$\mathbf{H} = \begin{bmatrix} \mathbf{A}^T & \mathbf{I}_{n-k} \end{bmatrix}$$

where $\mathbf{I}_{n-k}$ is the identity matrix of size $(n - k) \times (n - k)$.
Step 2: Form **G** of size $n \times k$ as follows:

$$\mathbf{G} = \begin{bmatrix} \mathbf{I}_k \\ \mathbf{A} \end{bmatrix}.$$

It is noted that $\mathbf{HG} = \mathbf{0}_{n-k,k}$ where $\mathbf{0}_{n-k,k}$ is the zero matrix of size $(n - k) \times k$. Note again $m = n - k$. The message **m** is encoded to produce a codeword **c** of size $n \times 1$ via $\mathbf{c} = \mathbf{Gm}$. Then, because of the definition of **G**, it is always seen that the result of multiplying **H** with **c** is the zero vector of size $m$, i.e.,

$$\mathbf{Hc} = \mathbf{HGm} = \mathbf{0}_m.$$

A decoder takes both the parity check matrix **H** and the corrupted word **r**, which is $\mathbf{r} = \mathbf{c} + \mathbf{e}$, where **e** is an error pattern. The decoder runs a message-passing algorithm [32] shown to be the standard decoding algorithm to remove **e**.

The principle behind the message-passing algorithm is to iteratively propagate probabilistic information among the variable and check nodes. The iterations are terminated if either the number of iterations exceeds a given number or an output is a codeword. Detailed explanations on how this algorithm operates are provided in [32], i.e., Algorithm 5.1 on page 220. The algorithm takes parameters such as **H**, **r**, maxIter, and $\varepsilon$, where maxIter is the number of maximum iterations, and $\varepsilon$ is the cross error probability that is used to determine the initial value of the algorithm.

The error-correction performance of the algorithm depends on both maxIter and the crossover error probability. If maxIter is small, the algorithm fails to obtain a converged solution. If it is large, the algorithm may take a considerably long computational time to obtain its solution. In the literature, maxIter is set from 10 to 20 in general. Next, the crossover error probability is set if the transition probability of a binary symmetric channel is either given or estimated. If this is improperly set, the decoding performance is degraded, leading to the poor error-correction capability. For the purpose in correcting errors, this parameter must be carefully considered.

The aim of using the decoder in ECCPoW, please make note of the fact that, is not to correct errors. Thus, there is no need to set maxIter and $\varepsilon$ strictly. One condition that we shall aim to satisfy is that all the miners in ECCPoW system have to use the same values for these parameters. This condition can be easily satisfied by letting them to be published fixed constants in the proposed implemented program. As the miners verify a newly published block before accepting it, there is no benefit not to follow and use different ones for these published parameters. That is, any proof obtained from arbitrary parameters other than the published ones must be rejected.

We will give details on how to construct the other parameters, such as **H** and **r**, in Section IV.

### B. FPGA and ASIC Implementation

LDPC decoders based on ASIC, a.k.a. ASIC-LDPC decoders are implemented to achieve low power consumption and fast processing. In the decoders, the check and variable nodes have to be physically connected using logical gates for a given parity check matrix. Fixed connections are to imply limited flexibility on the designs on the ASIC-LDPC decoders, making them only to support either a set of pre-defined parity check matrices or structured parity check matrices. We give our surveys related to existing ASIC-LDPC decoders as follows.

First, the ASIC-LDPC decoder in [23] supports quasi-cyclic parity check matrices decomposed into cyclic-shifted identity and zero matrices. These matrices have the same structure that is used in implementing the decoder. Second, the ASIC-LDPC decoder in [24] supports the parity check matrices included in the IEEE 802.16e system. These matrices are fixed; not change. Third, in [25], Hanzo *et al.*, reviewed the state-of-the-art of ASIC-LDPC decoders and stated that these decoders must take a bank of hardware to support many random parity check matrices. Namely, additional components such as memories, controllers and switchable interconnections are required, resulting in that these components occupy the most chip area in the decoders. They supported their statement by providing an example of [26] in which the ASIC-LDPC decoder supports about 100 parity check matrices, but its additional components occupy 75% of the total area of the decoder. This shows that there are no practical implementations on ASIC-LDPC decoders to support an infinite number of random parity check matrices.

There are FPGA-LDPC decoders that are the LDPC decoders implemented on FPGA chips. FPGA-LDPC decoders consume more power rather than ASIC-LPDC decoders do. But, it is much easier to reprogram the FPGA-LDPC decoders, which implies that they can achieve the more flexibility on the designs compared to ASIC-LDCP decoders. The FPGA-LDPC decoder in [27] supports parity check matrices up to $n = 65,000$. But, it is required to load a parity check matrix onto this decoder when it has to be changed, requiring additional time. In [28], Hanzo et al., stated that FPGA-LDPC decoders require additional routing and processing devices to support many parity check matrices. But, as they pointed, the use of these additional devices can lead to complex designs, increasing the cost of the decoders.

### IV. ERROR-CORRECTION CODES PROOF OF WORK

In this section, we give details on ECCPoW. For simplicity, we organize this section into four subsections. In the first subsection, we list fields belonging to a block header of ECCPoW and provide their simple explanations. In the second subsection, we illustrate an overall structure of ECCPoW and its explanations. In the third subsection, we explain how we construct two inputs that appear as we use a decoder of error-correction codes. In the last subsection, we give the definition of an ECC puzzle generation function and present how to define an ECC puzzle using this ECC function. We end the last subsection by giving routines for solving this ECC puzzle.

### A. Block header in ECCPoW

The block header of ECCPoW is defined to be a data structure that has eight fields such as timestamp, Merkle tree value, previous hash value, nonce, version, code length, row degree and column degree.

We use the fields such as version, timestamp, previous hash value, Merkle tree value and nonce to achieve in the purpose of guaranteeing the immutability property similar to Bitcoin.

We use the remains such as the code length, the row degree and the column degree to assign the size of hash vector and parity check matrix (PCM), which appears due to the usage of a decoder of a family of LDPC codes. As we will show in Section V, we change the difficulty level of a puzzle by varying the code length if the column degree and the row degree are fixed.

### B. Overall structure of ECCPoW

Fig. 1 is prepared to present an overall structure of ECCPoW. This structure consists of three parts such as *i*) the hash vector generation (HVG) part, *ii*) the LDPC decoder part and *iii*) the decision part. We explain each part as follows.

In the HVG part, we randomly generate a hash vector of size $n$ using a series of SHA256s taking the CBH with a given nonce generated by the nonce generator. The details how to generate this hash vector from the knowledge of the CBH will be given in the subsection IV.C.

In the LDPC decoder part, there is a decoder from a family of LDPC codes. This decoder takes the above hash vector and runs the message-passing algorithm [32] to yield a binary word **c**. It is noted that this decoder takes a parity check matrix (PCM) **H** determining the relation between an input and its corresponding output. The details how to construct this PCM will be given in the subsection IV.C.

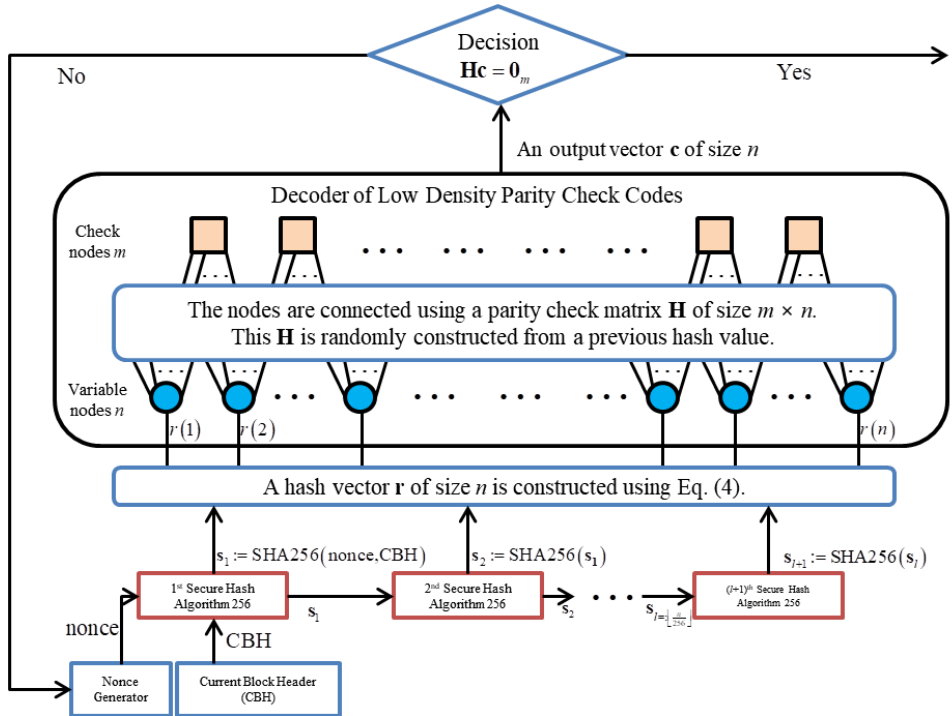

Fig. 1. An overall scheme of ECCPoW.

Last, in the decision part, the decision is made on the basis of the output provided by the decoder, as we have shown in Fig. 1.

*C. Construction of hash vector and parity check matrix in ECCPoW*

First, we provide the definition of hash vector **r** and give how we construct this hash vector using the current block header.

**Definition 1** – *Hash Vector*: A hash vector **r**, which is a vector of concatenating outputs of SHA256s, of size $n$ is defined to be:

$$\mathbf{r} := \begin{cases} \mathbf{s}_1[1:n] & \text{if } n \leq 256 \\ [\mathbf{s}_1 \cdots \mathbf{s}_l \ \mathbf{s}_{l+1}[1:j]] & \text{if } n > 256 \end{cases} \quad (4)$$

where $l = \lfloor n/256 \rfloor$, $j = n - 256 \times l$,

$$\mathbf{s}_1 := \text{SHA256}(\text{CBH}) \in \{0,1\}^{256} \quad (5)$$

and

$$\mathbf{s}_u := \text{SHA256}(\mathbf{s}_1) \in \{0,1\}^{256} \quad (6)$$

where $u = 2, 3, \ldots, l+1$ and CBH is the current block header.

The current block header represented to be CBH in (5) is the on-chain information that is stored in the Internet. Anyone thus can access this on-chain information, leading to that anyone can make the same hash vector that a miner made during his mining competition work.

Second, we give a construction method of PCM for a given previous hash. This construction method has to be designed to satisfy two conditions:

C1. Any verifier can reconstruct the PCM using on-chain information that the miner has used.

C2. Formation of a PCM can vary from block to block.

First, suppose that C1 is not met. One choice a miner can opt out is to include a constructed PCM in a block for making any verifier to check the validity of this block. This may result in, under the condition that the block size is fixed, reducing the number of transactions stored in the block. Second, suppose C2 is not met. We then remind the literature survey given in Section III. This survey is to indicate that developing ASIC-LDPC decoders for supporting a single PCM can be possible. Thus, if C2 is not satisfy, there is a possibility to develop ASIC mining devices. Thus, C1 and C2 must be satisfied simultaneously.

Now, we begin to explain the proposed construction method that can satisfy C1 and C2. First, we let you know that the PCM parameters such as the code length $n$, the row degree $w_r$ and the column degree $w_c$ are essentially required to construct a PCM **H**. More concretely, if they are provided, we can assign the value of $m$, the number of rows of **H** in (1). These parameters are included in the block header, as we have stated in the second paragraph in this section.

The proposed method is based on the method of Gallager [7]. This aims to construct **H** that can be decomposed into a set of sub-matrices of size $w_c \times n$ as follows:

$$\mathbf{H} = \begin{bmatrix} \mathbf{A} \\ \pi_1(\mathbf{A}) \\ \vdots \\ \pi_{w_c-1}(\mathbf{A}) \end{bmatrix} \in \{0,1\}^{\frac{nw_c}{w_r} \times n} \quad (7)$$

where $\pi_i(\mathbf{A})$ is the $i^{\text{th}}$ submatrix constructed by random per-mutation of the columns of $\mathbf{A}$, $\pi_i$ is the $i^{th}$ permutation order, and

$$\mathbf{A} := \begin{bmatrix} \mathbf{1}_{w_r} & & \\ & \ddots & \\ & & \mathbf{1}_{w_r} \end{bmatrix} \in \{0,1\}^{w_c \times n}$$

whose the $i^{th}$ row has $w_c$ 1s in a row from $(i-1) \times w_r$ to $i \times w_r$ and

$$\mathbf{1}_{w_r} := \begin{bmatrix} 1 & 1 & \cdots & 1 \end{bmatrix} \in \mathbf{1}^{1 \times w_r}.$$

Let $w_c$, $w_r$ and $n$ are fixed. Different PCM can be constructed by varying the permutation orders, leading to the serial creation of different PCMs. To come up with different random permutation from block to block, we use a previous hash. Namely, we let PBHV be assumed to be an array of 32 bytes to represent the given previous hash. We generate an initial seed value $S$ below

$$S := \text{PBHV}[0] + \text{PBHV}[1] + \cdots + \text{PBHV}[31] \quad (8)$$

where PBHV[$i$] is the $i^{th}$ element of PBHV. The first permutation order is generated using a seed value $S$. The $i^{th}$ permutation order is then generated using $S - i + 1$. The pseudo code of this construction method of PCM is given in Table VI.

We explain how the pseudo code given in Table VI can satisfy the conditions C1 and C2 mentioned earlier. In Step 4, the $i^{th}$ permutation order is constructed using $S - 1 + i$. Any verifier can get the same value $S$ without any communications because PBHVs are on-chain information and thus available within the chain. Thus, the verifiers can easily reconstruct what the miner has constructed in the past, which confirms that the proposed method satisfies C1. Second, PBHVs are hashes and thus possess the characteristics of random numbers; i.e., the initial seed value is a random number. All of the permutation orders are provided using the initial seed value. Thus, the orders vary from block to block, which confirms that the proposed method satisfies C2 as well.

### D. Construction of ECC puzzle generation functions and ECC puzzles

Let the current block header (CBH) except nonce be given. We define an ECC puzzle generation function by concatenating the HVG part and the decoder part.

**Definition 3** – *ECC puzzle generation function*: Let the current block header (CBH) except nonce be given. For this given CBH, an ECC puzzle generation function (ECCPGF) is defined to be a composite function as follows:

$$h_{\text{CBH}} : \{\text{nonce}\} \mapsto \mathbf{c} \in \{0,1\}^{n \times 1} \quad (9)$$

where $\mathbf{c}$ is the output of a decoder defined in Definition 4.

**Definition 4** – *Decoder*: A decoder takes both the hash vector $\mathbf{r}$ in (4) and the PCM $\mathbf{H}$ in (7) as its inputs and runs the message passing algorithm given in [32] to yield a vector $\mathbf{c}$ of size $n$:

$$\mathcal{D}_{MP} : \{\mathbf{r}, \mathbf{H}\} \mapsto \mathbf{c} \in \{0,1\}^{n \times 1}. \quad (10)$$

TABLE V. The pseudo codes for ECCPoW

| Inputs: CBH and PCM $\mathbf{H}$ | |
|---|---|
| Step 1: | A nonce is uniformly chosen from $[0, 2^{32} - 1]$ |
| Step 2: | Construct a HV $\mathbf{r}$ using (4) with a chosen nonce and the given CBH. |
| Step 3: | Obtain a vector $\mathbf{c}$ using (10) with the given PCM $\mathbf{H}$. |
| Step 4: | If the constraint in (11) is satisfied, then go to Step 5. |
| Step 5: | Block generation & broadcast |

TABLE VI. The pseudo codes to construct PCM

| Inputs: $n$, $w_c$, $w_r$ and BHV | |
|---|---|
| Output: $\mathbf{H}$ | |
| Step 1: | Construct $S$ using (8). |
| Step 2: | Construct $\mathbf{A}$ by following the statements below (7) and $\mathbf{H} = \mathbf{A}$. |
| Step 3: | for $i = 2$ to $w_c - 1$ |
| Step 4: | Construct $\pi_i$ with the seed value $S - i + 1$. |
| Step 5: | $\mathbf{H} = \begin{bmatrix} \mathbf{H} & \pi_i(\mathbf{A})^T \end{bmatrix}$. |
| Step 6: | end |
| Step 7: | $\mathbf{H} = \mathbf{H}^T$. |

When the CBH with a selected nonce is given, we can make a hash vector $\mathbf{r}$ according to (4). Using the previous hash value included in the CBH, we can make a PCM $\mathbf{H}$ as well. The decoder then takes both of them to yield the binary word, as we have shown in (10). This binary word is the output of ECCPGF, as we have shown in (9)

A mapping rule in the LDPC decoder depends on the form of $\mathbf{H}$. As we have discussed in the subsection IV.C, we can choose to vary $\mathbf{H}$ from block to block. As blocks are mined endlessly, infinitely many PCMs can be made. Thus, as we have mentioned in the subsection III.B, this can deter the development of an ASIC-LDPC decoder. Making ASIC chips to function as an ECCPGF becomes extremely difficult, if not impossible. Thus, we use this ECCPGF to define an ECC puzzle as follows.

**Definition 5** – *ECC Puzzle*: An ECC puzzle constructed using a given ECCPGF defined in (9) is defined to be

$$\text{find} \quad nonce \quad \text{subject to} \quad \mathbf{H} h_{\text{CBH}}(nonce) = \mathbf{0}_m. \quad (11)$$

Namely, this puzzle is a problem where we aim to find a nonce satisfying the constraint given in (11).

We shall note that this puzzle can be time-variant from block to block and resistant to ASIC chips. We will provide details in Section VI.

Last, we provide codes to solve a puzzle in Table V. In Step 1, a nonce is selected from 0 to $2^{32} - 1$. We construct a hash vector $\mathbf{r}$ using (4), as we have shown in Step 2. In Step 3, we execute the decoder to give an output by taking both $\mathbf{r}$ and $\mathbf{H}$. The decision is done using this output in Step 4. If the output is not a codeword, we repeat the routines from Step 1 to Step 4. Similar to Bitcoin, there is a case in which we cannot find the solution even we consider the whole nonces. In such a case, we modify the fields such as timestamp and Merkle tree value of the current block header. This modification can lead to the variation of the hash vector (4). We then repeat the whole routines with this modified block header.

## V. ANALYSIS ON ERROR-CORRECTION CODES PROOF OF WORK

To solve an ECC puzzle, we repeat the routines from Step 1 to Step 4 many times until finding a nonce that satisfies the constraint in (11). A simple question such that what is the number of trials for finding this nonce naturally arises. To this end, we conduct a probabilistic study for giving answers to the following three questions:

Q1. What is the number of hash cycles needed to solve a given puzzle?
Q2. Which are the parameters which affect the number of hash cycles needed?
Q3. How does the number of miners affect the number of hash cycles needed?

We define the hash cycle, the success event, and the mining game, respectively.

**Definition 6** – *Hash Cycle*: The single execution of the whole routines from Step 1 to Step 4 given in Table V is defined to be single hash cycle.

**Definition 7** – *Success Event*: A success event occurs if a nonce such that the decoder defined in (10) can return a codeword is found, i.e., the constraint in (11) is satisfied.

**Definition 8** – *Mining Game*: Let both a PCM $\mathbf{H}$ of size $m \times n$ and a CBH except for nonce be given. There are $M$ miners, and each use a single computer with the same computing capacity. A mining game (MG)

$$\mathrm{MG}\{\mathbf{H}, \mathrm{CBH}, M, p\} \tag{12}$$

is defined that the miners compete with each other in a race to hit the success event first. We let $p$ be the decoding success (DS) probability of the decoder for this given PCM $\mathbf{H}$, i.e.,

$$p := \Pr\{\mathbf{r} : \mathbf{Hc} = \mathbf{0}_m\} \tag{13}$$

where $\mathbf{c}$ is the output of the decoder defined in (10).

For a given PCM $\mathbf{H}$ of size $m \times n$, there are $2^k$ codewords:

$$\{\mathbf{c}_1, \mathbf{c}_2, \mathbf{c}_3, \cdots, \mathbf{c}_{2^k}\}.$$

Then, we define a sphere set for the given $i^{\mathrm{th}}$ codeword

$$\mathcal{A}(\mathbf{c}_i, l) \triangleq \{\mathbf{r} : \|\mathbf{r} - \mathbf{c}_i\|_h \le l \cap \mathbf{r} \in \{0,1\}^n\}$$

whose cardinality is

$$|\mathcal{A}(\mathbf{c}_i, l)| = \binom{n}{0} + \binom{n}{1} + \binom{n}{2} + \cdots + \binom{n}{l} = \sum_{l=0}^{s}\binom{n}{l}$$

where $s$ is a positive integer and $\|\mathbf{x}\|_h$ is the number of 1s in $\mathbf{x}$.

We assume that the decoder defined in (10) is optimal, implying that it can correct up to $t$ errors where $t$ is obtained by (3). The decoder always yields the $i^{\mathrm{th}}$ codeword when it takes an input belonging to the $i$th sphere, i.e.,

$$\mathcal{D}_{MP} : \{\mathbf{r}, \mathbf{H}\} \mapsto \mathbf{c}_i$$

where $\mathbf{r} \in \mathcal{A}(\mathbf{c}_i, t)$. Then, we have

$$\begin{aligned} p &= \sum_{i=1}^{2^k} \Pr\{\mathbf{c} = \mathbf{c}_i\} = \sum_{i=1}^{2^k}\sum_{l=0}^{t}\Pr\{\|\mathbf{r} - \mathbf{c}_i\|_h = l\} \\ &= \sum_{i=1}^{2^k}\Pr\{\mathbf{r} \in \mathcal{A}(\mathbf{c}_i, t)\} \end{aligned} \tag{14}$$

where $\mathbf{c}$ is an output of the decoder which takes a hash vector $\mathbf{r}$. Since the number of inputs that can be mapped into one of the codewords is

$$\sum_{i=1}^{2^k}|\mathcal{A}(\mathbf{c}_i, t)|,$$

the DS probability $p$ can be expressed as follows:

$$p = 2^{-n}\sum_{i=1}^{2^k}|\mathcal{A}(\mathbf{c}_i, t)| = 2^{k-n}|\mathcal{A}(\mathbf{c}_i, t)| = 2^{k-n}\sum_{l=0}^{\lfloor\frac{d-1}{2}\rfloor}\binom{n}{l} \tag{15}$$

where the third equality comes from (3).

Intuitively, the number of trials increases as $p$ decreases. It is natural to find which parameters make effects on $p$. To this end, we establish Proposition 1 to provide the behavior of $p$ in terms of the code length $n$ under the assumption that the row degree and the column degree are fixed.

**Proposition 1** – Let $w_c \ge 3$, $w_r > w_c$ be fixed constants and their ratio be a fixed constant as well

$$w_c/w_r =: \alpha \in (0,1). \tag{16}$$

Let the size of a given PCM $\mathbf{H}$ be $m \times n$. For any $0 < \delta < 1/2$, we have

$$2^{-n\alpha} \le p \le 2^{-n(\alpha - H(\delta/2))} \tag{17}$$

where $H(x)$ is the binary entropy function defined as follows:

$$H(x) = -x\log_2 x - (1-x)\log_2(1-x).$$

Indeed, let the ratio further satisfy the following:

$$\alpha \in (H(0.25), 1). \tag{18}$$

Then, the DS probability $p$ vanishes with increase in $n$.

**Proof**: From (1) with (16), we infer that

$$k - n = -n\alpha. \tag{19}$$

The results [7] state that the minimum distance of any regular LDPC code with constant $w_r$ and $w_c \ge 3$ can linearly increase with increase in $n$. This statement indicates that the distance can be expressed as for any $0 < \delta < 1/2$,

$$d = \lfloor \delta n \rfloor.$$

Substituting (19) into (15) leads to an upper bound to $p$

$$p \le 2^{k-n}\sum_{l=0}^{\lfloor\frac{\delta n}{2}\rfloor}\binom{n}{l} \le 2^{k-n}2^{nH\left(\frac{\delta n}{2n}\right)} \le 2^{-n(\alpha - H(\delta/2))} \tag{20}$$

where the second inequality comes from the fact that any integers $n \geq k \geq 1$ with $k/n \leq 0.5$

$$\sum_{l=0}^{k} \binom{n}{l} \leq 2^{nH(k/n)}.$$

By assuming $t = 0$ in (14), the lower bound is obtained below:

$$p = 2^{k-n} |\mathcal{A}(\mathbf{c}_i, t)| \geq 2^{k-n} |\mathcal{A}(\mathbf{c}_i, 0)| = 2^{k-n} = 2^{-n\alpha}.$$

Last, let the ratio satisfy (18). Then, the term of the exponent in the upper bound in (17) is negative because for any $0 < \delta < 1/2$,

$$H(\delta/2) < H(0.25).$$

Thus, increasing $n$ makes the upper bound on $p$ reduced. This leads to that $p$ vanishes as $n$ goes to infinity. ∎

We provide quick discussions on Proposition 1. First, for the fixed constant $\alpha$, decreasing the code length $n$ makes the lower bound on the DS probability $p$ grow. Second, if we select any pairs of $w_c$ and $w_r$ which can satisfy (18), then $p$ vanishes as $n$ increases. This means that no one solves an ECC puzzle using these pairs for sufficiently large $n$. We thus have to avoid selecting such pairs for preventing this critical problem.

We define a random variable to represent the number of hash cycles required to end a MG and provide its statistical properties, respectively.

**Definition 8** – *Random Variable*: For a given MG, $X_M$ is defined as a random variable that represents the number of hash cycles to end this given MG, where the subscript $M$ denotes the number of miners forging simultaneously and independently. We call this random variable First Success Hash Cycle (FSHC).

**Theorem 2** – Let an MG{$\mathbf{H}$, CBH, $M$, $p$} be given. Then the distribution that FSHC occurs at the $l^{\text{th}}$ hash cycle is

$$\Pr\{X_M = l\} = p_{\text{f,a}}^{l-1}(1 - p_{\text{f,a}})$$

which is a geometric distribution with a parameter

$$(1 - p_{\text{f,a}}) \tag{21}$$

where $p_{\text{f,a}} := (1-p)^M$ is the probability that all of the miners fail to succeed in solving a given puzzle. Then, we have

$$\mathbb{E}[X_M] = (1 - p_{\text{f,a}})^{-1} \tag{22}$$

and

$$\mathbb{V}[X_M] = p_{\text{f,a}}(1 - p_{\text{f,a}})^{-2}.$$

**Proof**: The proof is clear, as $X_M$ follows the geometric distribution with (21). We thus omit it. ∎

For the constant $\alpha$ defined in (16), Proposition 1 shows that the DS probability $p$ grows with decrease in the code length $n$. When $p$ grows, the parameter (21) reduces and converges to a real positive number. Thus, the expected value gets reduced and converges to a number, implying that an ECC puzzle becomes easier as $n$ gets smaller.

TABLE VII. The lower bound given in (23)

| $w_c = 4$ and $w_r = 5$ | $M = 1$ | $M = 5$ | $M = 20$ |
|---|---|---|---|
| Lower bounds, i.e., $\delta_1 = 0.3238$ in (23) | | | |
| $n = 80$, $k = 12$ | $1.58 \times 10^4$ | $0.31 \times 10^4$ | $0.08 \times 10^4$ |
| $n = 120$, $k = 24$ | $6.03 \times 10^7$ | $1.20 \times 10^7$ | $0.30 \times 10^7$ |
| $n = 160$, $k = 32$ | $2.46 \times 10^9$ | $0.49 \times 10^9$ | $0.12 \times 10^9$ |

Next, we consider a case in which $n$ grows. In this case, the upper bound given in (17) is too loose to be considered unless $\alpha$ satisfies (18). Thus, we require another upper bound on $p$ which is tighter than the previous upper bound. To this end, we invoke a table of [31]. For a certain pair of the column degree $w_c$ and the row degree $w_r$, this table was obtained and given in the form of an upper bound $\delta_1$ to the relative minimum distance $\delta_0$, the ratio of the minimum distance $d$ to the code length $n$. For $w_c = 4$ and $w_r = 5$, for example, the upper bound is 0.3238. For $w_c = 4$ and $w_r = 8$, it is 0.1765. The result was obtained from an asymptotic analysis; i.e., by letting $n$ go to infinity.

We use the upper bound $\delta_1$ to obtain a tighter bound on the DS probability $p$ as follows:

$$p \leq g(n, k, \delta_1) \tag{22}$$

where $\delta_1$ is given in the table of [31] for a certain pair of $w_c$ and $w_r$ and

$$g(n, k, \delta) := 2^{k-n} \sum_{l=0}^{\left\lfloor \frac{\lfloor n\delta \rfloor - 1}{2} \right\rfloor} \binom{n}{l}.$$

The bound in (22) is obtained by simply replacing the minimum distance with the upper bound, respectively:

$$d := n\delta_0 \leq \lfloor n\delta_1 \rfloor.$$

Once the upper bound on $p$ has been obtained, we can use it to find a lower bound to the expected value of FSHC below:

$$\frac{1}{1 - (1 - g(n, k, \delta_1))^M} \leq \mathbb{E}[X_M] \tag{23}$$

We can examine the behavior of (22) with respect to the code length $n$ and the number of miners $M$. In Table VII, we provide the lower bounds to the expected value by varying either $n$ or $M$. They are obtained for a case in which the column degree $w_c$ and the row degree $w_r$ are 4 and 5, respectively. We can see that the lower bound increases with increase in $n$ for the fixed $M$. That is, it increases from $1.58 \times 10^4$ to $2.46 \times 10^4$ when $n$ is increased from 80 to 160. This result implies that increasing $n$ makes an ECC puzzle more difficult to solve. For the other pairs of $w_c$ and $w_r$ given in [31], the same result is observed.

Now, for the fixed code length $n$, we consider the behavior of the expected value of FSHC by growing the number of miners $M$. Intuitively, as more miners are involved in solving a given puzzle, this puzzle has to end early. In addition, if an infinite number of miners work, any MG has to end at the $1^{\text{st}}$ hash cycle. These intuitions can be confirmed by Corollary 1 given below.

**Corollary 1**: Let a MG{**H**, CBH, $M$, $p$} be given. The expected value of a FSHC given in (22) decreases with increase in the number of miners $M$. In particular, this value can converge to 1 as $M$ goes to infinity.

**Proof**: It is immediately seen that

$$\frac{d\mathbb{E}[\mathrm{X}_M]}{dM} = -\frac{\log p_{\mathrm{f,a}}}{p_{\mathrm{f,a}}} \leq 0$$

implying that the expected value given in (22) is a decreasing function of $M$. As $M$ goes to infinity, the parameter defined in (21) goes to one. Thus, the expected value converges to one. ∎

The decoding process, i.e., Step 3 in Table V, can occupy the most computational time in a single hash cycle. In this decoding process, matrix-vector products are required, implying that the computational cost to run a single hash cycle can be modeled as $O(mn)$. Each miner uses the same single computer in a given MG. Thus, we can assume that each miner only runs $\tau$ operations per second. This assumption makes us define an expected value of a block generation time as follows.

**Definition 10** – *Block Generation Time*: A MG{**H**, CBH, $M$, $p$} is given. Each miner is assumed to run $\tau$ operations per second. Then, the block generation time $T$ can be defined as

$$T \coloneqq \tau^{-1}\mathbb{E}[\mathrm{X}_M]O(mn). \tag{24}$$

Both Proposition 1 and Theorem 2 can indicate that the expected value in (24) is an increasing function of the code length $n$. We thus immediately conclude that the block generation time $T$ is an increase function of $n$.

## VI. DISCUSSIONS ON ECCPoW

We prepare this section to provide plentiful discussions on ECCPoW. To this end, we begin to define general properties of PoWs and explain how ECCPoW can have them. We introduce a new property that ECCPoW can only have. This new property makes ECCPoW become a solution to the problems such as M1 and M2 which we have stated in Section I.

We now begin to define properties below.

P1. A puzzle has to be time-consuming, but it is easy to check whether a given solution is correct or not.
P2. Any previous solution cannot be used to find a current solution.
P3. A puzzle can be solved with overwhelming probability 1 if and only if miners follow the routines of PoW.
P4. The difficulty of a puzzle can change.
P5. A puzzle can be time-variant from block to block.

The existing PoWs have the properties from P1 to P4. Let us begin to consider how bitcoin satisfies them. First, each puzzle is expected to be solved per 10 minutes. In contrast, validating a given solution can be instantly done. Second, SHA256 takes the block header to get its hash. The block header in Bitcoin has the timestamp field. Due to this timestamp field, the contexts of the current block header can be different to those of any previously mined blocks. Thus, any solutions given in these mined blocks are useless in finding a current solution. Thus, P2 holds. Third, the number of possible hashes is $2^{256}$ while the number of solutions is about $2^{(256-L)}$ where $L$ is a pre-defined value according to the difficulty of the provided block header. Thus, a possibility that a provided nonce is a solution is $2^L$. In the 567,657th block of bitcoin, $L$ is 72. This is to imply that for this block, the probability that a randomly given nonce can be a solution is $2^{-72}$. Thus, all the miners have to follow the routines in Table I to solve a puzzle, meaning P3. Next, whenever 2016 blocks are mined, the difficulty periodically changes, implying P4. Last, existing PoW does not hold P5. The reasons for this issue will be given after explaining how ECCPoW satisfies P5.

Now, we begin to explain how ECCPoW can satisfy all the properties using the results established in the previous section.

**Corollary 2**: ECCPoW can satisfy the first property P1.

**Proof**: For a given nonce, we complete the verification whether this nonce is a solution or not by conducting the steps from Step 1 to Step 4 in Table V. This verification requires a single construction of a hash vector and a single execution of the decoder (10). In contrast, to solve an ECC puzzle, one has to do work by repeating them many times, as we have stated in Section V. ∎

**Corollary 3**: ECCPoW can satisfy the second property P2.

**Proof**: As we have stated in Definition 5, we construct an ECC puzzle using an ECCPGF in Definition 3. This ECCPGF takes a hash vector as its input and produces a binary word as its output. The decision is made based on this binary word. To make this hash vector, we put a given set of the block header with a nonce through SHA256. Similar to Bitcoin block header, our block header includes the timestamp field. Thereby, one cannot effortlessly create a particular hash vector even if one has the full knowledge of all the previous solutions, i.e., all the collection of the nonces each of which was a solution to mine a block in the past. This makes ECCPoW satisfy P2. ∎

Let us assume that there exists a knowledgeable miner who can solve ECC puzzles by referring to an input-output mapping table of ECCPoW without actually carrying out the decoding work. This miner can then mine blocks much faster than other honest miners can. To show the non-existence of such a miner in ECCPoW, we prepare Corollary 4. This corollary is to imply that all miners cannot but have to run the decoder to solve ECC puzzles.

**Corollary 4**: ECCPoW satisfies the third property P3.

**Proof**: Such a malicious miner can appear under an assumption that this miner has a mapping table that maps a given input to its corresponding output without actually carrying out the decoding work (10). That is, he puts each hash vector into the decoder to find what this decoder returns. He then uses this known information to construct the mapping table. However, the number of hash vectors is $2^n$. For sufficiently large $n$, i.e. $2^{256} \sim 10^{77}$ for $n = 256$, we can easily see that this construction is impossible. Besides, this table depends on a PCM **H**, meaning that the mapping table is updated newly whenever **H** varies. As we have stated in the subsection IV.C, **H** can be set to vary from block to block. Thereby, whenever he aims to solve a new ECC puzzle, he has to construct a new mapping table. The above assumption there exists the mapping table, we thus conclude, is invalid. Thereby, any miners in ECCPoW cannot but have to obey the routines in Table V to solve puzzles; this leads to the conclusion that ECCPoW satisfies P3. ∎

Now, we remind the discussions regarding Theorem 2. In the discussions, we have proven that an ECC puzzle can become easier as the code length $n$ gets smaller. We then have empirically shown that the increase in $n$ makes this puzzle more difficult to solve. They are to indicate that ECCPoW holds P4.

**Corollary 5**: ECCPoW can satisfy the fourth property P4.
**Proof**: It is clear; we thus omit it. ∎

A PCM **H** defines a mapping function of the decoder in (10). This decoder is used to define an ECCPGF in (9) which is used to construct an ECC puzzle. We conclude that this ECC puzzle can be a function of **H**. As we have mentioned in the subsection IV.C, we construct **H** using a previous hash, i.e., the $i^{th}$ PCM is made using a hash of the $(i–1)^{th}$ block. Such a construction can make **H** time-variant from block to block, leading to that the ECC puzzle can also be time-variant from block to block. This discussion leads to a conclusion that ECCPoW holds P5 below.

**Corollary 6**: ECCPoW can satisfy the fifth property P5.
**Proof**: It is clear; we thus omit it. ∎

We now prepare to provide discussions how ECCPoW can be a solution to the problems such as *i*) M1 the re-centralization of the mining markets and *ii*) M2 the huge energy consumption for mining. As miners mine new blocks continuously, an infinite number of PCMs are constructed. Hence, we cannot count how many the number of PCMs is required in advance. Also, we cannot expect what PCMs will be generated. We remind the example given in [25], showing that there is no ASIC-LDPC decoder to support the infinite number of PCMs. By combining this example with the fact that there is an infinite number of PCMs are required, we thus conclude that the LDPC decoder in ECCPoW is operated by either graphical processing units or CPUs. This is the fundamental reason that ECCPoW becomes the solution remediating the aforementioned problems M1 and M2 which we have stated in Section I. This is the most remarkable contribution of ECCPoW.

We end this section by providing a reason that SHA based PoW does not hold P5. We consider SHA256 as a puzzle generation function (PGF) in Bitcoin because it is used to generate a puzzle. This PGF is fixed regardless of the height of a block that miners aim to mine. The puzzle made using a fixed PGF is not time-variant at all, leading to that such a puzzle cannot hold P5.

## VII. CONCLUSIONS

PoW is fundamental to public blockchains, as it can be used to prohibit an unauthorized modification of mined blocks. For existing PoWs, ASIC mining devices have been introduced and used to mine blocks. The usage of such devices can cause the problems such as M1 and M2 that we have stated in Section I.

In this paper, as a solution to these problems, we proposed a new proof-of-work using error-correction codes which we call ECCPoW. To the best of our knowledge, this is the first study in which a decoder of LDPC codes is applied to the consensus part of blockchain. Specifically, we combined this decoder with SHA256 to construct a composite function named as ECCPGF (9). We used ECCPGF to define a corresponding ECC puzzle (11) and provided the routines to solve a given ECC puzzle.

We also studied the behavior of the expected value of the number of hash cycles for solving a given ECC puzzle. We showed that this value can be either increased or decreased as varying the code length, the size of a hash vector taken by the decoder, and the number of miners. Indeed, we discussed how ECCPoW can satisfy the five properties defined in Section VI, which shows the value of ECCPoW as a general PoW.

As we have reviewed in Section III, there is no ASIC decoder to support an infinite number of LDPC codes. By motivated this survey, we intended to vary the codes from block to block. As a result, we made ECCPGF time-variant, meaning that its mapping function can vary from block to block. This leads to the time-variant property P5 defined in Section VI. This is the most innovative part of ECCPoW in repressing the advent of ASICs, implying that the problems caused from the usage of ASICs can be solved using our ECCPoW.

We have implemented ECCPoW and forked two blockchains such as Bitcoin and Ethereum by replacing their consensus with the implemented ECCPoW. We name these forked versions as BTC-ECC and ETH-ECC, respectively. All of the source codes including ECCPoW, BTC-ECC and ETH-ECC can be available in a GitHub site [34]. We have also provided manuals that shows how to install them, how to compile them and how to run. We believe that anyone can easily operate their own BTC-ECC or ETH-ECC by following these manuals. We believe that this site can be the first repository that makes people in the error-correction codes community get involved in the blockchain community.